**Synthesis of $Zn_2NbN_3$ ternary nitride semiconductor with wurtzite-derived crystal structure**

Andriy Zakutayev
National Renewable Energy Laboratory, Golden CO 80401 USA
E-mail: andriy.zakutayev@nrel.gov

**Abstract**

Binary III-N nitride semiconductors with wurtzite crystal structure such as GaN and AlN have been long used in many practical applications ranging from optoelectronic to telecommunication. The structurally related $ZnGeN_2$ or $ZnSnN_2$ derived from the parent binary compounds by cation mutation (elemental substitution) have recently attracted attention, but such ternary nitride materials are mostly limited to II-IV-$N_2$ compositions. This paper reports on synthesis and characterization of zinc niobium nitride $Zn_2NbN_3$ – a previously unreported $II_2$-V-$N_3$ ternary nitride semiconductor. The $Zn_2NbN_3$ thin films are synthesized using a single-step adsorption-controlled growth, and a two-step deposition/annealing method that prevents loss of Zn and N. Measurements indicate that $Zn_2NbN_3$ crystalizes in wurtzite-derived structure, in contrast to chemically related rocksalt-derived $Mg_2NbN_3$ compound synthesized here for comparison using the two-step method. The estimated wurtzite lattice parameters are a = 3.36A and c = 5.26A, (c/a = 1.55), and the optical absorption onset is at 2.1 eV for this cation-disordered $Zn_2NbN_3$. For comparison, published computational studies predict cation-ordered $Zn_2NbN_3$ to be a semiconductor with effective wurtzite c/a = 1.62 and a band gap of 3.5 - 3.6 eV. Overall, this work expands the wurtzite family of nitride semiconductors, and suggests that other ternary nitrides should be possible to synthesize.

**Introduction**

Nitride materials (nitrides) is an interesting class of compounds with many useful properties. Nitrides with wurtzite (WZ) crystal structure such as GaN have been long used for optoelectronic applications, for example in (In,Ga)N alloys for light emitting diodes (LEDs)[1], and in (Al,Ga)N alloys for radio-frequency (RF) high electron mobility transistors (HEMTs)[2], due to their direct and wide band gaps. Rocksalt (RS) structured nitride semiconductors such as ScN have narrower band gaps that are often indirect, and attracted attention as thermoelectric materials[3] and for its p-type doping [4]. Alloying WZ and RS nitride semiconductors in WZ-(Al,Sc)N alloys with strong piezoelectric response[5] has been used for film bulk acoustic resonators (FBARs), and has recently led to emergence of strong ferroelectric response in this class of materials[6] sparking interest for nonvolatile memory applications. These examples illustrate that more nitride semiconductors, and their alloys, with diverse compositions and unique properties are needed to advance future device applications.

One approach to increase the diversity of nitride semiconductors is using "cation mutation" principle, widely used in other material chemistries.[7][8] In this approach, a formula unit of a binary semiconductor in doubled or tripled, and two of the same cations are substituted for two different ones with the same total charge to satisfy octet rule, leading to a new ternary compound with the same crystal structure. For example, in the Zn-IV-$N_2$ family of materials[9], ternary $ZnGeN_2$ with a well-matched lattice constant and a narrower band gap than GaN [10] attracted interest for LEDs applications, fueled by theoretically prediction of large band gap tuning at a fixed lattice constant by disorder [11]. As other examples $ZnGeGa_2N_4$ (a pseudo-binary alloy between $ZnGeN_2$ and GaN[12]) has been synthesized by metal–organic chemical vapor deposition (MOCVD)[13], ternary $LiSi_2N_3$ and $LiGe_2N_3$ compounds have been synthesized[14], an Li-containing quaternary nitride semiconductors have been predicted for ultraviolet optoelectronics[15], all in wurtzite-derived crystal structures. Mg-based ternary nitride semiconductors with RS-derived crystal structure, have been recently synthesized[16] with highly tunable electron density [17] and large electron mobility[18], and high predicted dielectric constants [16].

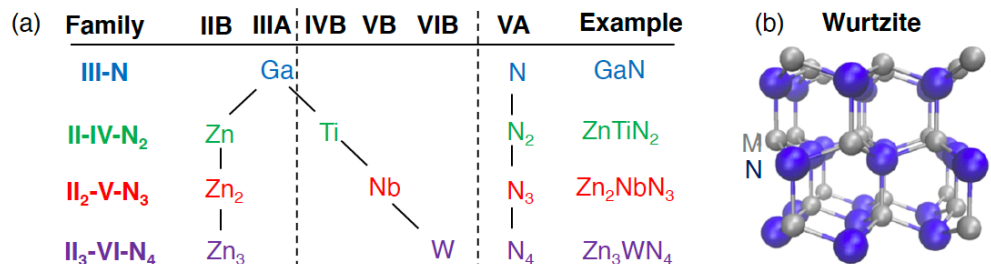

**Figure 1.** Composition and structure of ternary nitride materials. (a) Cation mutation diagram for several families ternary nitride semiconductors, including II-IV-$N_2$ (e.g. $ZnTiN_2$), $II_2$-V-$N_3$ (e.g. $Zn_2NbN_3$ studied here), and $II_3$-VI-$N_4$ (e.g. $Zn_3WN_4$), with wurtzite III-N (e.g. GaN) being parent compounds. (b) A wurtzite-derived crystal structure with 4-fold coordinate metal and nitrogen atoms for $Zn_2NbN_3$ synthesized here and for other related ternary nitrides.

**Figure** 1 shows a cation mutation approach to derive Zn-based ternary nitride semiconductors with wurtzite crystal structure, which is more general compared to basic III-V → II-IV-$N_2$

substitution. In this II-IV-$N_2$ → $II_2$-V-$N_3$ → $II_3$-VI-$N_4$ ternary series, the content of the group-II metal is increased, while transition etal is substituted by another one with higher valence state (IV→V→ VI). The series as a whole can be described by the $II_{(G-3)}$-TM-$N_{(G-2)}$ formula [16] where G is the group of the high-valence transition metal TM or a corresponding main group element (e.g. Sb). Several members of these ternary semiconductor families have been recently synthesized and characterized, such as WZ-$Zn_3MoN_4$[19], WZ-$Zn_3WN_4$[20], WZ-$Zn_2SbN_3$[21], but others such as WZ-$Zn_2NbN_3$ zinc niobium nitride have not been experimentally reported to date.

The Zn-based cation-mutated ternary nitride compounds are challenging to synthesize due to lower melting point and higher vapor pressure of Zn and $Zn_3N_2$ compare to other metals. As the synthesis temperature increases Zn and N tend to sublime from the sample, leaving it rich in the other component. For example, substantial loss of Zn has been observed in the 300-400C temperature range during thin film synthesis of $ZnGeN_2$[22][23] and $ZnSnN_2$[24][25]. This is one of the reasons that high-pressure is often used for bulk synthesis of Zn-containing ternary nitrides. For instance $ZnSnN_2$[26], $CaZn_2N_2$[27] and $Zn_2PN_3$[28] have been synthesized at the pressures as high as 5 - 10 GPa. Thus, new synthesis methods that can suppress the loss of Zn and N would be beneficial for synthesis of Zn-based ternary nitride materials.

This manuscript reports on the first successful synthesis of $Zn_2NbN_3$ – a new ternary nitride semiconductor. Stoichiometric $Zn_2NbN_3$ thin films were synthesized by one-step absorption-limited growth in overflux of Zn and N, and by a two-step approach of co-sputtering at ambient temperature followed by rapid thermal annealing at elevated temperature, designed to prevent volatilization of Zn species. The resulting $Zn_2NbN_3$ material crystallizes in wurtzite-derived crystal structure with random Zn and Nb occupancy of the cation sites, estimated wurtzite lattice parameters ratio of 1.55, and the optical absorption onset at 2.1 eV. These experimental results are qualitatively consistent with published computational predictions for cation-ordered $Zn_2NbN_3$, but the theoretical c/a = 1.62 and $E_g$ = 3.5-3.6 eV band gap are higher than the corresponding experimental values. These results indicate that WZ-$Zn_2NbN_3$ is a new ternary semiconductor promising for ferroelectric or optoelectronic applications, and suggest it should be possible to synthesize other Zn-based cation-mutated ternary nitride semiconductors using similar methods.

**Experimental Results**

$Zn_2NbN_3$ thin films were synthesized using a two-step approach, to address the Zn volatility challenge. First, Zn-Nb-N thin film precursors of 150 nm thickness are deposited by co-sputtering at ambient temperature, leading to complete Zn incorporation and atomically dispersed mixing of the constituent elements. Second, these atomically-dispersed precursors are subject to rapid thermal annealing (RTA) at high temperature for short time, to crystallize the material without losing much Zn. Metal composition and phase content of the samples have been monitored by x-ray fluorescence (XRF) and x-ray diffraction (XRD) after each annealing experiment. In addition, two AlN diffusion barriers of 50 nm thickness each are used to prevent side reactions with Si substrate during RTA and with ambient atmosphere during characterization. For comparison, $Zn_2NbN_3$ films without the AlN capping layer, $Zn_2NbN_3$ on heated $SiO2$ substrate, as well as reference $Mg_2NbN_3$ and $Zn_3N_2$ films are also deposited using similar synthesis conditions. More details about these synthesis and characterization experiments are provided in the Methods section.

Chemical composition of metals in various Zn-Nb-N samples as a function of deposition or annealing temperature (3 min annealing time) is shown in **Figure 2a.** As annealing temperature of AlN-capped $Zn_2NbN_3$ samples increases (Fig.2a), composition and thickness remain constant up to 650 - 750C, which is higher than 550-650C for uncapped $Zn_2NbN_3$ films, and much higher compared to 300-400C temperature limit during thin film deposition of $Zn_2NbN_3$. Above each of these critical temperature ranges, the $Zn_2NbN_3$ decomposes into solid RS-NbN loosing Zn and N. The critical temperatures for annealing $Zn_2NbN_3$ samples correlate well with the rapid loss of thickness from $Zn_3N_2$ films deposited as a reference (550C and 650C for uncapped and capped samples respectively). The 650C anneal for 3 min maintains the composition of capped $Zn_2NbN_3$ samples, but longer annealing times also led to the loss of Zn-containing species (see **Figure S1** in Supporting Information). However, the loss is more gradual with time in contrast to the quick loss with temperature, with some variation in time depending on the thickness and microstructure of the AlN capping layer.

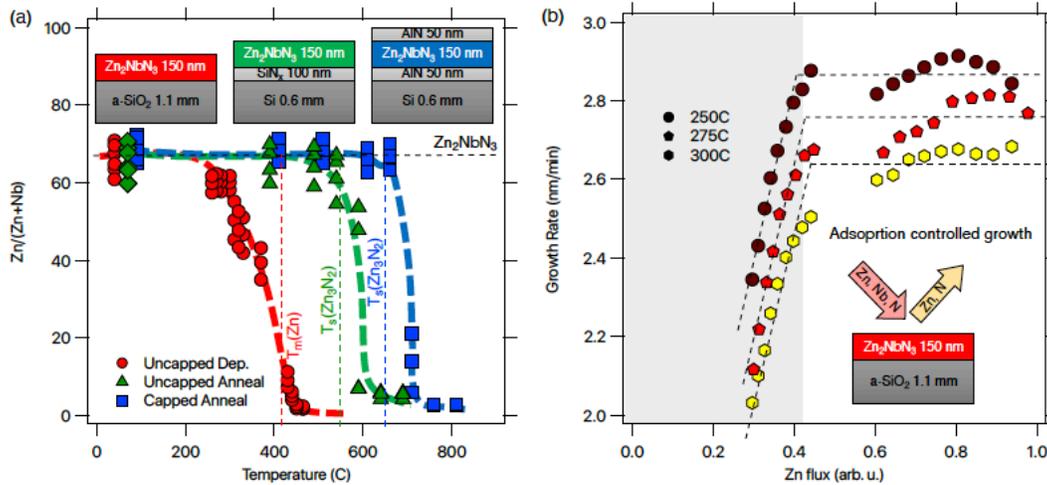

**Figure 2** $Zn_2NbN_3$ experimental synthesis. (a) Measured cation composition of Zn-Nb-N samples as a function of annealing or deposition temperature, with the stoichiometric $Zn_2NbN_3$ composition, sublimation temperature of Zn3N2, and the melting temperature of Zn marked by dashed lines. Inset shows a schematic illustration of three different Zn-Nb-N sample stacks (b) Growth rate of Zn-Nb-N samples as a function of relative Zn flux for three different temperatures, with dashed lines as guides to the eye. Inset schematically illustrates the adsorption-controlled growth regime where the rate is independent of flux.

The preferential loss of $Zn_3N_2$ at elevated synthesis or annealing temperatures points to the possibility of adsorption-controlled growth of $Zn_2NbN_3$ thin films, where the composition of the product is tolerant to fluctuations in the chemical potential or the flux of the precursors. **Figure 2b** shows growth rate for $Zn_2NbN_3$ films on a-$SiO_2$ substrate at 3 different temperatures as a function of Zn flux estimated from ambient temperature deposition. The curves steeply increase for low Zn flux and saturate above the critical flux indicating desorption of volatile $Zn_3N_2$ species. As the synthesis temperatures increases, the critical flux increases and the growth rate decreases, both as expected. Thus the growth of stoichiometric $Zn_2NbN_3$ can performed slightly above $Zn_3N_2$ evaporation temperature, so the rate is limited by dynamic equilibrium between adsorbing and desorbing volatile species.[29] Such self-limiting adsorption-controlled growth mode is new to ternary nitrides, but it has been used for controlling composition of many ternary oxides[30] by molecular beam epitaxy (MBE), and some ternary chalcogenides [31] by physical vapor deposition (PVD). The adsorption-controlled growth of $Zn_2NbN_3$ would useful in tuning defect properties for future practical application.

Structural measurement of of stoichiometric $Zn_2NbN_3$ samples are presented in **Figure 3a**. The measured XRD patterns of $Zn_2NbN_3$ samples annealed at 650 C for 3 min are consistent with wurtzite-derived crystal structure of polycrystalline Zn2NbN3, where Zn and Nb cations occupy cation sites in random fashion. The $Zn_2NbN_3$ samples deposited at 400C in adsorption-controlled mode show strong preferential orientation with c-axis perpendicular to the a-SiO2 substrate plane, as indicated by the absence of all other XRD peaks. The wurtzite-derived (WZ) crystal structure of $Zn_2NbN_3$ material is clearly different from the rocksalt-derived (RS) crystal structure of related $Mg_2NbN_3$ material synthesized here for comparison using the same 2-step approach (Fig.3a).

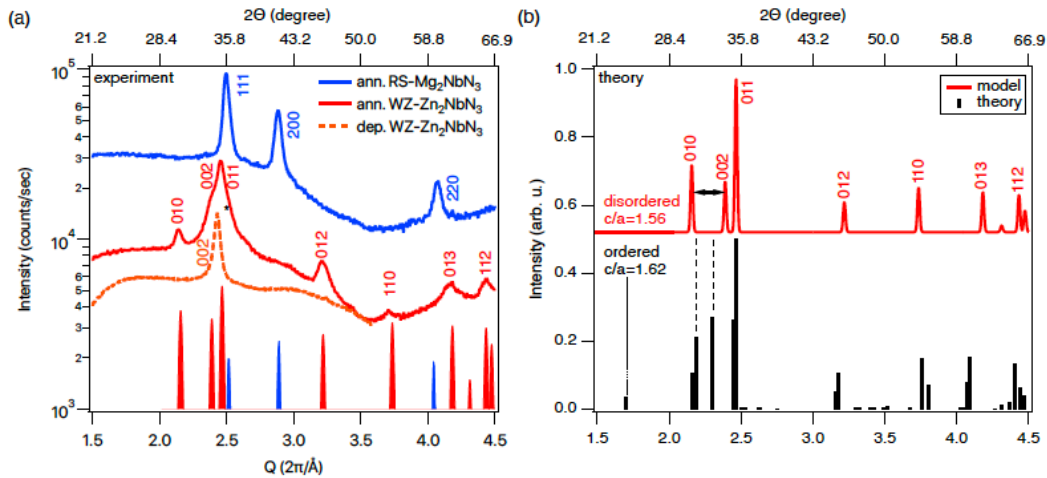

**Figure 3**. $Zn_2NbN_3$ crystal structure. (a) Measured XRD pattern of WZ-$Zn_2NbN_3$ samples annealed under AlN cap and as-deposited on heated substrate, in comparison with RS- $Mg_2NbN_3$ annealed under AlN cap. Asterisk corresponds to the expected position of the major AlN peak (b) Modeled XRD patterns of $Zn_2NbN_3$, for cation-ordered structure with close to ideal wurtzite c/a=1.62 in comparison with measured cation-disordered structure with wurtzite c/a=1.55

Results of XRD measurements (Fig.3a) show that for the annealed $Zn_2NbN_3$ samples, the wurtzite 002 peak at Q=2.39 $A^{-1}$ (2Θ=34.1) appears very close to the 101 peak at Q=2.46$A^{-1}$ (2Θ=35.2). The corresponding wurtzite lattice constants determined from (100) and (101) peaks are a=3.36A and c=5.26A, indicating the lattice ratio of c/a=1.55. As shown in **Figure 3b**, this measured c/a=1.55 value for the cation-disordered wurtzite $Zn_2NbN_3$ is lower than c/a=1.62 for the theoretically predicted cation-ordered wurtzite $Zn_2NbN_3$ which is close to expected √8/3=1.63 for the ideal wurtzite structure. No evidence of cation ordering is observed in the synthesized $Zn_2NbN_3$ in Fig.3b, as indicated by absence of XRD peak at Q=1.7 $A^{-1}$ (2Θ=24.1) and other minor reflections at higher angle. The for $Zn_2NbN_3$ structure consistent with experimental XRD measurements in provided in Table S1 and Table S2 of Supporting Information, and in an attached CIF file

Measured optical properties of $Zn_2NbN_3$ are shown in **Figure 4a.** Strong optical absorption of $Zn_2NbN_3$ films grown at 300C starts above 2.1 eV, corresponding to a decrease in transmission below 600 nm photon wavelength for 150 nm thin films (Fig.4a inset). The absorption coefficient reaches $10^5$ cm$^{-1}$ close to 3 eV. Below the 2.1 eV optical absorption threshold, the $Zn_2NbN_3$ films show a shallower exponential tail of optical absorption, which in prior literature on other ternary nitrides has been attributed to of sub-gap tail states due to cation disorder on wurtzite lattice. Four-point probe sheet resistance measurements in plane of slightly Zn-rich 150 nm thick WZ-$Zn_2NbN_3$ films indicate that they are highly resistive (> 1 GOhm measurement limit of the instrument), as expected for undoped >2 eV band gap material. The resistivity decreases below the 1 GOhm measurement limit for slightly Zn-poor WZ-$Zn_2NbN_3$ films where reduced Nb valence state or lower-band gap impurities with RS structure may be expected.

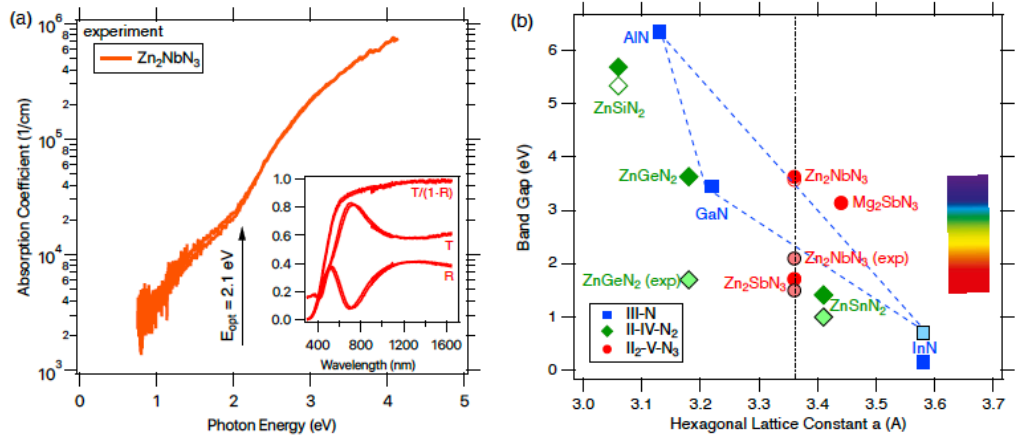

**Figure 4**. Measured and calculated optical properties of $Zn_2NbN_3$. (a) Optical absorption spectra of WZ-$Zn_2NbN_3$ synthesized by adsorption-controlled growth at 300C. Inset shows transmittance and reflectance spectra use to obtain the absorption spectrum. (b) Summary plot of measured and calculated band gaps and structural parameters for $Zn_2NbN_3$ material, other wurtzite-derived $II_2$-V-$N_3$ compounds, and several related III-N binary and II-IV-$N_2$ ternary nitrides. Insets show a visible spectral range.

**Computational discussion**

Experimental $Zn_2NbN_3$ structure and property results presented above are in qualitative agreement with theoretical predictions, but there are important quantitative differences. A summary of the measured properties of $Zn_2NbN_3$ is presented in **Table 1**, and compared with calculated properties this and related ternary and binary nitride materials. These property predictions are publicly available in several papers [27] [32] and in computational materials databases such as materials.nrel.gov[33] and materialsproject.org.[34]

Theoretically, $Zn_2NbN_3$ is predicted to crystalize in a cation-ordered wurtzite structure similar to that of $Zn_2PN_3$ which has orthorhpombic $Cmc2_1$ symmetry (space group 36) [28] [27] [32]. The predicted $Zn_2NbN_3$ lattice constants in this orthorhombic structure are $a_0$=9.852A, $b_0$=5.782A, $c_0$=5.386A [27], which corresponds to wurtzite lattice parameters of a = 3.34A = $b_0/\sqrt{3} \approx a_0/3$ and c=5.39A, close to idea c/a = $\sqrt{8/3}$ = 1.63 for wurtzite material. For the calculated c/a = 1.62 lattice parameters, the maximum predicted piezoelectric coefficient is 0.687 C/m$^2$ along the b-axis and bulk modulus is 139 GPa [34] all comparable to other known III-N (GaN) and II-IV-N$^2$ (e.g. ZnGeN$_2$) materials listed in Table 1. However, the experimental results indicate much smaller c/a = 1.55 (>3% different from the predictions), which is comparable to that reported in Al$_{0.7}$Sc$_{0.3}$N alloys with strong piezoelectric [5] and ferroelectric [6] properties. The measured low c/a suggests potential use of Zn$_2$NbN$_3$ in these applications.

**Table 1.** Summary of measured and calculated structural, optoelectronic (materials.nrel.gov), and piezoelectric (materialsproject.org) properties of Zn$_2$NbN$_3$, in comparison with calculated properties for related II$_2$-V-N$_3$, II-IV-N$_2$, and III-V materials with hexagonal wurtzite-derived crystal structure.

| Material | Hex. a, A | Hex. c, A | Hex. c/a | Eg eV | m*e, m0 | m*h, m0 | Piezo axis | Piezo Mod., C/m2 | Bulk Mod., Gpa |
|---|---|---|---|---|---|---|---|---|---|
| **Zn$_2$NbN$_3$ (exp)** | 3.36 | 5.26 | 1.55 | 2.1 | | | | | |
| **Zn$_2$NbN$_3$** | 3.36 | 5.44 | 1.62 | 3.43 | 1.08 | 4.39 | a | 0.687 | 139 |
| **Zn$_2$SbN$_3$** | 3.36 | 5.47 | 1.63 | 1.71 | 0.37 | 2.46 | c | 0.745 | 132 |
| **Mg$_2$SbN$_3$** | 3.44 | 5.42 | 1.58 | 3.14 | 0.44 | 2.38 | c | 1.416 | 115 |
| **ZnSiN$_2$** | 3.06 | 5.07 | 1.66 | 5.34 | 0.54 | 4.29 | a | 1.845 | |
| **ZnGeN$_2$** | 3.18 | 5.25 | 1.65 | 3.63 | 0.49 | 2.42 | a | 0.594 | 163 |
| **ZnSnN$_2$** | 3.41 | 5.54 | 1.62 | 1.41 | 0.43 | 2.13 | c | 1.070 | 133 |
| **AlN** | 3.13 | 5.02 | 1.60 | 6.35 | 0.53 | 2.07 | c | 1.676 | 195 |
| **GaN** | 3.22 | 5.24 | 1.63 | 3.46 | 0.48 | 2.47 | c | 0.619 | 172 |
| **InN** | 3.58 | 5.79 | 1.62 | 0.16 | 0.39 | 2.12 | a | 5.551 | 106 |

The calculated band gaps for cation-ordered Zn$_2$NbN$_3$ are nearly direct, around 3.1 eV at HSE level and 3.5 eV at G$_0$W$_0$ level[27], consistent with computational data available at materials.nrel.gov. Experimentally measured optical absorption onset is around 2.1 eV, smaller than computational predictions due to cation disorder. The predicted effective masses are quite low, m$_h$*=0.99m$_0$ for holes (2x smaller than GaN) and m$_e$*=0.39m$_0$ for electrons (2x larger than GaN) [27], both lower than the corresponding density of states masses available at materials.nrel.gov. The calculated formation enthalpy is H$_f$=-0.585 eV/at and decomposition energy (energy below convex hull) is -0.122 eV/at,[32] similar to thermochemistry predictions at materials.nrel.gov. Future experiments would be required to validate the computation predictions for formation enthalpy and effective masses.

The combination of structural and optoelectronic properties of Zn$_2$NbN$_3$ makes it an interesting compound for designing epitaxial heterostructures with low strain using other II$_2$-V-N$_3$ materials. **Figure** 4b shows a plot of calculated band gaps versus hexagonal wurtzite lattice constants for Zn$_2$NbN$_3$, as well as other materials synthesized recently with wurtzite-derived crystal structure and 2-1-3 stoichiometry such as Zn$_2$SbN$_3$ [21] and Mg$_2$SbN$_3$[35]. The calculated band gap and lattice constant of Zn$_2$NbN$_3$ fall right in the miscibility gap of the (Al,In)N alloy tie-line, interesting for ultraviolet optoelectronic applications. The calculated 3.5 eV band gap of Zn$_2$NbN$_3$ is significantly larger than the calculated 1.7 eV band gap of Zn$_2$SbN$_3$, while their hexagonal lattice constant different by only 0.2%. Experimentally measured optical band gaps of these two lattice-matched materials show smaller difference (2.1 eV vs 1.5 eV), but still in a relevant range to (In,Ga)N alloys used in visible light-emitting diodes. Thus the resulting disordered or ordered $Zn_2(Nb,Sb)N_3$ alloys and their low-strain superlattices may address the miscibility gap problems of (In,Ga)N or (Al,In)N alloys and lattice mismatch issues with InN/GaN or AlN/InN heterostructures respectively, in design of light emitting diodes and lasers with nitride materials

It is interesting that the alloys between $Zn_2NbN_3$ and $Zn_2SbN_3$ materials of $II_2$-V-$N_3$ family are predicted to show a much large change in the band gap for a much smaller change in the lattice constant, compared to that III-N or II-IV-$N_2$ families of alloys. Unless it is a computational error, this large change may originate from the difference in the electronic structure (atom-projected density of states). Such large difference may be caused by the presence of transition metal (e.g. Nb $d_0$-states) versus main group (e.g. Sb $s_0$-states) elements in these $II_2$-V-$N_3$ compounds. However, this difference appears to be not as strong in experimentally known main group III-N (GaN vs. InN) or II-IV-$N_2$ ($ZnGeN_2$ vs. $ZnSnN_2$) compounds, or their computationally predicted transition metal relatives (e.g. WZ-$ZnGeN_2$ vs $ZnTiN_2$[36]). So the exact origin of the large change in the band gap for a smaller change in the lattice constant in these $II_2$-V-$N_3$ materials remains to be fully understood.

**Outlook**

Successful synthesis of $Zn_2NbN_3$ points to experimental opportunities with other ternary nitrides. Another predicted $II_2$-V-$N_3$ compound $Zn_2TaN_3$, with similar structure and properties [27][32], should be possible to synthesize using the one-step or two-step methods reported here. Such analogy in synthesis of wurtzite-derived ternary nitrides with 2$^{nd}$ vs 3$^{rd}$ row transition metals has been demonstrated for group-VI elements $Zn_3MoN_4$[19] vs $Zn_3WN_4$[20]. On the other hand, group-IV transition metals from 2$^{nd}$ and 3$^{rd}$ rows (Zr and Hf) together with Zn are predicted to form ternary nitrides in a different layered crystal structure distinct from wurtzite[27][32][36]. However, WZ-$ZnTiN_2$ would be a good synthesis candidate for expansion of wurtzite family of ternary nitrides to group-IV transition metals. These potential future experiments would lead to further expansion of wurtzite family of ternary nitrides and facilitate their future applications in optoelectronic and other devices.

Another interesting research direction is to attempt synthesis of the corresponding $Mg_2NbN_3$ compounds (Fig.2) in wurtzite-derived crystal structure. This material has been predicted and synthesized in rocksalt-derived structure[16], but the polymorph energy difference with h-BN like structure related to wurtzite is <10 meV/at according to calculated formation enthalpies available at materials.nrel.gov. Experimentally stabilizing Mg in tetrahedral coordination has been shown feasible in the recently synthesized $Mg_2SbN_3$ compound with WZ-derived structure [35], and would be of interest to Mg-ion multi-valent battery applications[37]. One possible approach to stabilize the WZ- $Mg_2NbN_3$ structure is by alloying with WZ-$Zn_2NbN_3$ reported here, similar to Mg/Zn alloys with RS/WZ structural transition in binary (Zn,Mg)O [38]. Because of the small polymorph energy difference, only minute amount of Zn alloying or a thin Zn-rich seed layer may be sufficient to tip this balance and stabilize WZ-related polymorph of $Mg_2NbN_3$. Similar control of polymorphism by small amount of alloying agent has been recently reported for zinc-stabilized manganese telluride (Zn,Mn)Te alloys with WZ structure[38] or WZ-MnTe on ultrathin ZnTe seed layers[39], and is well known for cubic yttria-stabilized zirconia (YSZ) [40].

**Summary and Conclusions**

In summary, $Zn_2NbN_3$ has been synthesized for the first time, by adsorption-controlled growth on heated substrate in overflux of Zn and N, and by a two-step process designed to prevent Zn and N loss. The two-step process consists of of depositing thin film precursors at ambient temperature and annealing them at elevated temperature, with or without AlN capping layer. Experimental measurements indicate that $Zn_2NbN_3$ crystallizes in cation-disordered WZ-derived crystal structure with 2.1 eV optical band gap and c/a = 5.26A/3.36A = 1.55 lattice constants, distinct from the RS-derived structure of $Mg_2NbN_3$. These experimental results for cation-disordered $Zn_2NbN_3$ are consistent with previously published computational predictions for cation-ordered $Zn_2NbN_3$, but the calculated band gap (3.6 eV) and lattice constant ratio (c/a = 1.62) are larger than experimental measurements.

In conclusion, it should be possible to alloy WZ-$Zn_2NbN_3$ reported here with WZ-$Zn_2SbN_3$ synthesized earlier, to achieve large tuning of the band gap for a negligible ~0.2% change in the lattice constant attractive for heteroepitaxial integration in optoelectronic devices. Additionally, the $(Zn,Mg)_2NbN_3$ alloys with WZ structure may be a feasible approach to stabilize Mg in 4-fold coordination for applications in divalent batteries and/or ferroelectric devices. Overall, the experimental WZ-$Zn_2NbN_3$ results presented in this paper suggested that other new $II_2$-V-$N_3$ and $II_3$-V-$N_4$ materials and their alloys with exciting properties are awaiting to be synthesized beyond of the well-explored II-IV-$N_2$ family of cation-mutated ternary nitride semiconductors.

**Methods**

For the first step of the two-step synthesis approach, Zn-Nb-N thin films have been deposited by co-sputtering from Zn and Nb targets in N plasma on Si substrates without intentional heating. The 50 mm diameter targets were excited by radio-frequency (RF) field of Zn=30W and Nb=60W, and the N plasma intensity was enhanced by RF plasma source of N=350W. The deposition pressure was held at $8x10^{-3}$ Bar by flowing 3 sccm of $N_2$ and 6 sccm of Ar through a custom vacuum chamber with base pressure of $<10^{-9}$ bar, with lower effective base pressure in the plasma zone enabled by a cryoshroud cooled by liquid nitrogen. The 60 min deposition resulted in a 150 nm thin Zn-Nb-N films with some variation of composition across the sample. Similar approach has been used for deposition of $Zn_2NbN_3$ films on heated glass (a-$SiO_2$) substrate in adsorption-controlled self-limiting growth mode, we well as $Mg_2NbN_3$ and $Zn_3N_2$ thin films synthesized here for comparison.

For the second step of the two-step synthesis approach, the films deposited at ambient temperature were heated using a rapid thermal annealing (RTA) furnace (Ulvac-5000) in flowing $N_2$ at ambient pressure. The annealing temperature and time were varied in the T=400-800C range for 3 min, and t=3-12 min range at 650C, with each subsequent anneal performed on the previous sample but at higher temperature and longer time. The heat up time was held constant at 1 min, and a natural cool down rate was used (2-3 min to down 100C). Prior to the anneal, the furnace was purged with N2 for at least 3 min at 100C to remove residual $O_2$ and $H_2O$ vapor. Before and after some of the ambient-temperature Zn-Nb-N deposition, a diffusion barrier and a capping layer were deposited, both consisting of a 50 nm thick AlN sputtered from Al target held at 60W with conditions otherwise equal to $Zn_2NbN_3$.

The thin film samples were characterized for composition using x-ray fluorescence (XRF), for crystal structure using x-ray diffraction (XRD), for optical properties using transmission/reflection spectroscopy, and for electrical properties using four-point probe sheet resistance measurements. The XRF measurements of metals were performed in air using Fischer XDV-SDD instrument with Rh x-ray source and Si drift detector. Quantification of nitrogen would require other methods beyond the scope of this work. The XRD patterns were collected in Cu Ka radiation on Bruker D8 diffractometer equipped with a 2D detector. The absorption spectra were obtained from transmittance and reflectance measurements with a custom fiber optics spectrometer. Electrical measurements were attempted on a custom four-point probe instrument with W probes with 1 mm spacing. The characterization data were analyzed and correlated with deposition data using custom CombIgor[41] software package for Igor Pro, available for download through www.combigor.com. The resulting data will be made publicly available through High Throughput Experimental Materials (HTEM) database[42] at www.htem.nrel.gov.

**Acknowledgements**
This work was authored at the National Renewable Energy Laboratory, operated by Alliance for Sustainable Energy, LLC, for the U.S. Department of Energy (DOE) under Contract No. DE-AC36-08GO28308. Funding provided by Office of Science (SC), Office of Basic Energy Sciences (BES), Materials Chemistry program, as a part of the Early Career Award "Kinetic Synthesis of Metastable Nitrides". Drs. Keisuke Yazava, Paul Todd, Kevin Talley, and Brooks M. Tellekamp are gratefully acknowledged for their feedback to an early version of this manuscript. The views expressed in this article do not necessarily represent the views of the DOE or the U.S. Government.

**ORCID IDs**
Andriy Zakutayev  https://orcid.org/0000-0002-3054-5525

**Supporting Information:**

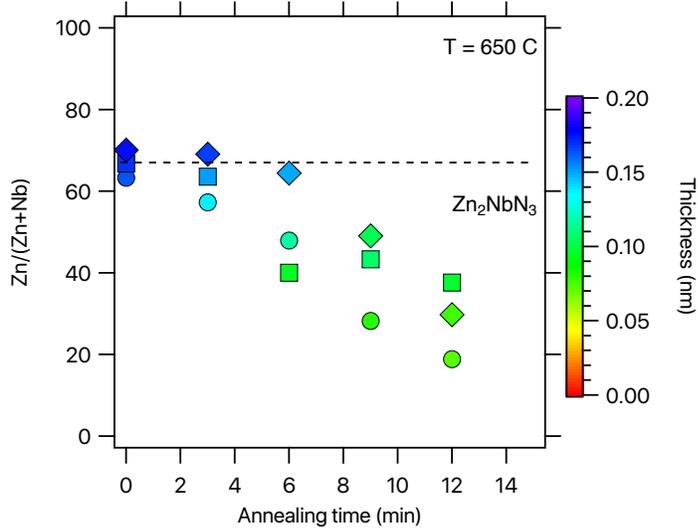

**Figure S1**. Measured cation composition of Zn-Nb-N samples as a function of annealing time for 650C annealing temperature, with the stoichiometric $Zn_2NbN_3$ composition marked by dashed line.

**Table S1** Unit cell of $Zn_2NbN_3$ determined from experiments

| | |
|---|---|
| Chemical Formula: | $Zn_2NbN_3$ |
| Z: | 1 |
| Space Group: | $P\,6_3\,m\,c$ |
| | *Polar* |
| Crystal System: | Hexagonal |
| a: | 3.3600 Å |
| c: | 5.2600 Å |
| Cell Volume: | 51.427 Å$^3$ |
| Asymmetric Unit: | 3 sites |
| Unit Cell: | 6 sites/unit cell |
| | 0.0778 atoms/Å$^3$ |
| Density: | 5.7132 g/cm$^3$ |

**Table S2** Atom positions of $Zn_2NbN_3$ determined from experiments

| Site | Occ. | x/a | y/b | z/c | m |
|---|---|---|---|---|---|
| M1 | $Zn_{0.67}$ | 0.3333 | 0.6667 | 0.3812 | 2 |
| M1 | $Nb_{0.33}$ | 0.3333 | 0.6667 | 0.3812 | 2 |
| N1 | $N_{1.0}$ | 0.3333 | 0.6667 | 0.0000 | 2 |